          [2004/06/22 1.2 (KOF); 2001/04/25 1.1 (PWD)]

\documentclass[a4paper,twocolumn]{Gaia2004} 
\usepackage{times}      
\usepackage{epsfig}     
\usepackage{natbib}     

\def\aabun{[$\alpha$/Fe]}
\def\extinct{{\bf A}($\lambda)$}
\def\av{A$_{\rm V}$}
\def\teff{${\rm T}_{\rm eff}$}
\def\logg{$\log g$}
\def\feh{[Fe/H]}

\def\parallax{$\pi$}

\def\vv{{\bf v}}

\def\rv{{\bf r}}

\title{Object classification and the determination of stellar parameters}

\author{C.A.L.\ Bailer-Jones}
\affil{Max-Planck-Institut f\"ur Astronomie, K\"onigstuhl 17, 69117
Heidelberg, Germany}

\bibpunct{(}{)}{;}{a}{}{,}  

\begin{document}


\maketitle

\begin{abstract}
Gaia will observe more than one billion objects brighter than $G=20$,
including stars, asteroids, galaxies and quasars. As Gaia performs
real time detection (i.e. without an input catalogue) the intrinsic
properties of most of these objects will not be known a priori.  An
integral part of the Gaia data processing is therefore to classify
everything observed. This will be based primarily on multiband
photometry provided by Gaia, but should also make optimal use of the
high resolution spectroscopy (for brighter stars) and the parallaxes.
In addition to a broad classification, we can also determine
fundamental stellar parameters, in particular effective
temperature, metallicity and the line-of-sight interstellar
extinction. Such information will be essential for fully exploiting the
astrometric part of the Gaia catalogue for stellar population studies.
However, extracting this information is a 
significant challenge, and will need to make use of appropriate
multidimensional data analysis techniques. I outline some of the problems
and the strategies being developed to tackle them. 
\end{abstract}

\section{Introduction}
The general astrometric principle of Gaia is similar to that of
Hipparcos. It scans the sky with a pre-defined scanning law measuring
the relative positions of objects. However, the two missions are
fundamentally different in various ways, one of the most relevant
being that Gaia goes much deeper, namely to about $G$\,=\,20
rather than $H_p$\,=\,12.4 as with Hipparcos.  Thus whereas
Hipparcos had an input catalogue, Gaia performs real-time onboard
detection. Consequently, we generally have very little prior
information about the targets which Gaia observes.

For this reason, Gaia is equipped with multiband CCD photometry in
order to characterize its targets.  As described elsewhere
in these proceedings, this consists of 4--6 broad band filters in the
Astrometric instrument, and around 12 medium band filters in the
Spectroscopic instrument.

The main scientific goal of Gaia is to study the composition, origin
and evolution of our Galaxy. Its main contribution to this topic is
establishing very accurate stellar distances and 2D (or 3D) space motions,
enabling us to study the 3D spatial structure and 3D kinematic phase
space of different types of stars in the Galaxy. However, such
information is of limited use if it cannot be associated with the
intrinsic physical properties of these stars, in particular their
abundances, masses and ages (or evolutionary state).  Hence Gaia is
not ``just'' about producing a catalogue of highly accurate
astrometry and multiband photometry on hundreds of millions of stars.
An integral part of the mission and data processing is to add
essential scientific value to these by providing fundamental physical
information on the targets.  The challenge is
to design a classification system -- and develop appropriate
algorithms -- which can take the heterogeneous Gaia data and extract
reliable estimations of the classes of objects and their
physical parameters.  This article gives a brief outline of the goals,
requirements and issues facing this work.

\section{Overall requirements and\\available data}

The main objectives of the classification are as follows
\vspace*{-2ex}
\begin{description}
\item[Discrete Source Classification]Determination of whether an object is a star, galaxy, quasar or asteroid
etc. This could also include the use of morphological information.
\item[Estimation of Astrophysical Parameters (APs)]For those objects identified as stars, determine their intrinsic
physical properties. The relevant (and obtainable) ones are effective
temperature, \teff, surface gravity, \logg, metallicity, \feh, and
line-of-sight interstellar extinction, \av. Although this last one is
of course not intrinsic to the star, we would ideally determine it on
a star-by-star basis, so we can consider it as such.
Other APs of interest (and which could be determined for bright stars
with the spectroscopy from the RVS instrument) include: alpha-process
elements, \aabun, CNO abundance anomalies, the microturbulence
velocity, rotational velocity  and activity.
\item[Identification of unresolved binaries]Most stars are in multiple systems. Some of these can be recognised
from the astrometry, and a few will be visual binaries, but most will
go undetected in this way. Nonetheless, with favourable brightness
ratios, a binary could be detected from the shape of its composite
spectral energy distribution. This is important for determining the
stellar mass function (as opposed to the system mass function) and for
investigating the evolution of stellar clusters.
\item[Identification of new types of objects]The history of astronomical discovery shows that new instruments,
surveys and data analysis techniques lead to new, unexpected
discoveries.  Thus with Gaia we must be open to the prospect of
detecting new types of objects (summarized by the cliche ``expect the
unexpected''.) This includes new types of variable stars, rare stars
(e.g.\ brief phases of stellar evolution), abnormal abundance patterns
or multiple systems. Those supervised classification methods which are
commonly used for determining stellar parameters from spectra are
generally forced to classify new types of objects into pre-existing
classes. New objects would therefore go undetected (and samples of
known types of objects would be contaminated). Thus special attention,
including the use of unsupervised methods, is required to deal with
this.
\end{description}

These classification tasks must rely mostly on the photometric data,
as only these extend to the magnitude limit of Gaia's onboard
detection and thus astrometry. But recall that the photometry is
obtained in two separate instruments, BBP and MBP.  Only the former is
obtained at the same spatial resolution as the astrometric data. Thus
in sufficiently crowded fields we will only have BBP data, that is
between 4 and 6 broad photometric bands.\footnote{We may still have
the MBP data -- if it is deemed worth transmitting to the ground --
but only then of composite objects. This might be
useable.} For bright stars, say brighter than $V=15$, we will also have
reasonable quality RVS spectra, which will add considerable
information, in particular on detailed abundances, peculiarities,
rapid rotation and so forth. Finally, the parallaxes will of course
also be valuable, as discussed in the next section.

\section{Stellar astrophysical parameter estimation}

The most fundamental properties of a star are its mass, age and
chemical composition.  Of course, age is not directly observable and
masses can only be determined directly (i.e.\ dynamically) in select
binary systems.  Thus we must rely on indirect atmospheric indicators
which can be obtained from the spectral energy distribution (SED).
In particular, we are interested in the effective temperature, \teff,
surface gravity, \logg\ and the iron-peak metallicity, \feh.  Combined
with the parallax and interstellar extinction, the luminosity, radius
and mass can be determined. 

\begin{table*}[t]
\begin{center}
\begin{minipage}{12cm}
\begin{tabular}{llll}
\hline 
\multicolumn{4}{l}{{\it non-astrometric parametrizer:}} \\
nSED, (RVS)	& $\Rightarrow$	& \teff, \logg, \feh, & \\ 
	& 	&  \extinct, BC, \aabun ?	& atmospheric model 	\\ 
\\
\multicolumn{4}{l}{{\it additional use of astrometry gives:}} \\
SED, BC, \parallax, \extinct\	& $\Rightarrow$	& L	& $2.5\log L - f(\rm{SED,BC})$\\
&&& \hspace*{1em}$= A - 5\log{\pi}$ \\ 
L, \teff\	& $\Rightarrow$	& R	& $L = 4 \pi R^2 \sigma T_{\rm eff}^4$ \\
\logg, R	& $\Rightarrow$	& M	& $g=GM/R^2$	\\
\\
SED, RVS, \vv(t), \rv(t)	& $\Rightarrow$	& detect unresolved binaries	& orbital model \\
SED(t), RVS(t)	 		& $\Rightarrow$	& detect variables		& variability model \\
\hline
\end{tabular}
\caption[]{Stellar parameters derivable from the Gaia data.
SED=spectral energy distribution (ca.\ 15 photometric measures in medium
and broad band filters); nSED=normalized SED (absolute flux
information removed); RVS=radial velocity spectrum; BC=bolometric
correction; \parallax=parallax; \extinct=interstellar extinction
function; \vv(t) \& \rv(t)=point source velocity and position as a
function of time (from c.\ 85 observations over five years).}
\label{parameters}
\end{minipage}
\end{center}
\end{table*}

Most work on stellar parametrization has relied on relatively high resolution
spectra from which \teff, \logg\ and \feh\ have been determined. Gaia
is rather different in that it observes at lower spectral resolution
but measures absolute fluxes as well as parallaxes.
Table~\ref{parameters} shows how stellar parameters can in principle
be derived from these data.  The distance measurement accuracy for
V\,=\,15 is 1\% at 1\,kpc and 5\% at 5\,kpc. At V\,=\,18 these are
about 4\% and 20\% respectively. (These improve by a factor of two or
more for late-type or very reddened stars.) Thus some 20 million stars
will have their distances determined to better than 1\% and have high
precision SEDs. If \teff\ can be established to 1\% then the radii of
many of these stars is determinable to within 2\%. If \logg\ can be measured to
0.2~dex, then provided R (radius) can be established to within 10\%, a mass
determination to within 50\% is possible without calibration from
binary systems. Although poor for an individual star, it becomes
statistically meaningful for a large sample of similar stars, which is
where Gaia's strength lies. Better individual masses will be possible
from calibration using the tens of thousands of visual binaries
observed by Gaia for which masses should be obtained to within 10\%
(and many thousands within 1\%). Individual ages (possibly with large
uncertainties) can be quantified from evolutionary models.

A proper treatment of interstellar extinction is very important.
Without accurate line-of-sight extinction measurements, the accurate
parallaxes and apparent magnitude measurements cannot be converted
into absolute magnitudes and thus intrinsic luminosities and radii.
For example, to determine the radius to 2\%, the extinction must be
measured to within 0.03 mags.

When trying to determine several astrometric parameters from a dataset
there exists the problem of parameter degeneracy, i.e.\ two different
astrophysical parameters manifesting themselves in the same way in the
SED in certain parts of the astrophysical parameter space. An example
is \teff\ and extinction in late-type stars, where lowering \teff\ has
a similar effect on the SED (at low resolution) as increasing the
extinction.\footnote{The radial velocity spectrum will help for the brighter
stars as this reddening-free information provides an independent
measure of the stellar parameters.} Clearly, for degenerate cases, a
parametrization algorithm is required which can give a range of
possible parameters, and not just a single set.

Most stellar systems consist of more than one component. Undetected
binaries bias the parameter determinations when the brightness ratio
is small (e.g.\ a higher luminosity for a given \teff\ leads to an
erroneous \feh\ determination).  Many long period and/or distant
binaries will go undetected with the astrometry.  In these cases,
parametrization techniques are required which can identify binary
stars from their composite SEDs and ideally parametrize both
components.

\section{The approach to classification and regression}

\begin{figure}[ht]
\begin{center}
\epsfxsize=0.45\textwidth \epsfbox{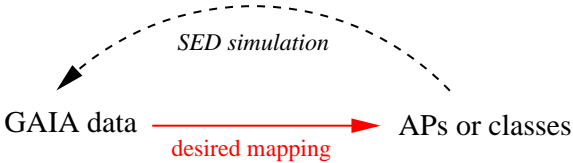}
\end{center}
\caption{Classification or parametrization is the process of
determining the mapping from a data domain to a class or astrophysical
parameter (AP) domain. The opposite mapping is equivalent to the
simulation of the data, e.g.\ the emergent stellar spectral energy
distribution (SED).}
\label{mapping}
\end{figure}

\begin{figure}[t]
\begin{center}
\epsfxsize=0.35\textwidth \epsfbox{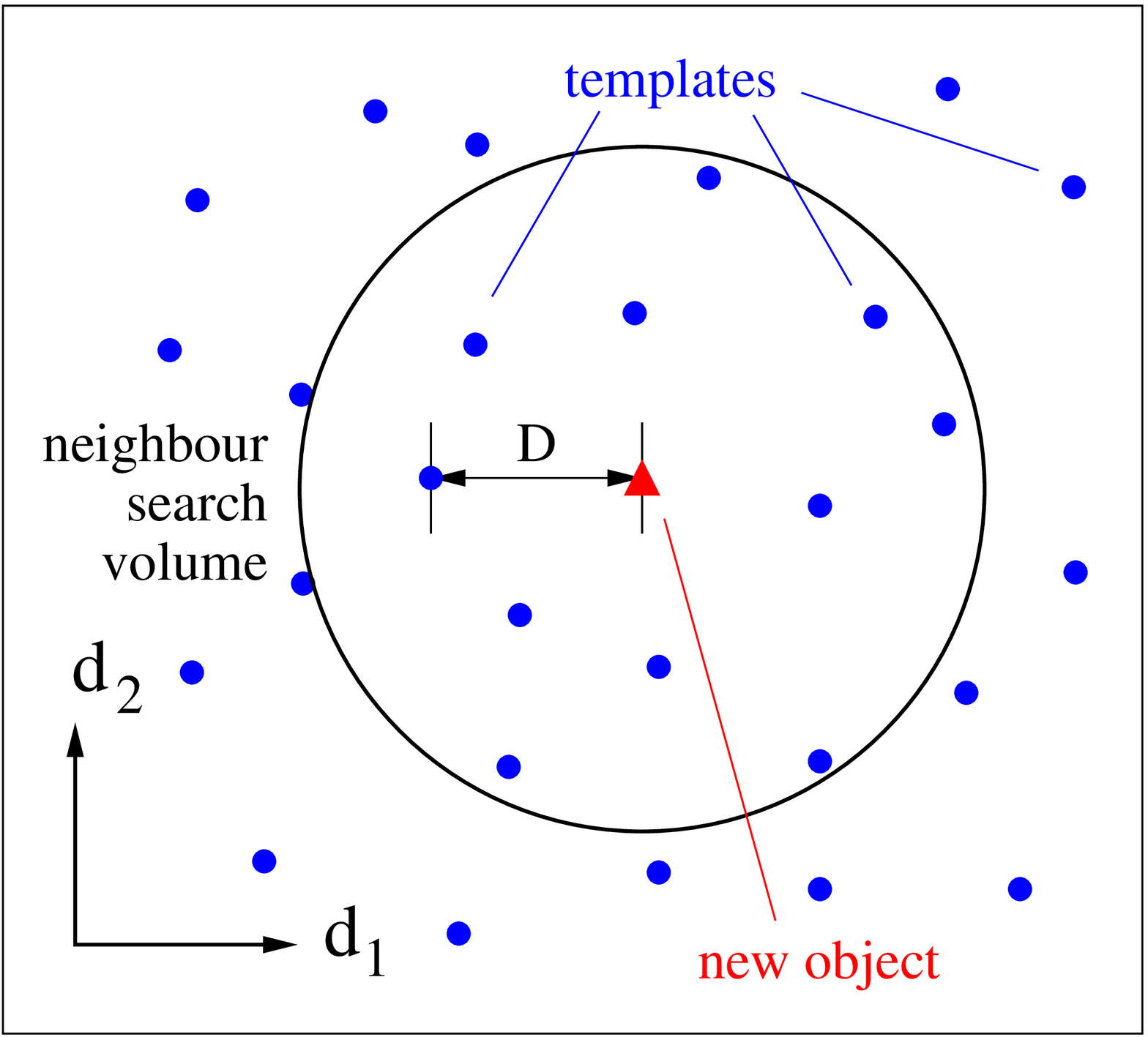} 
\epsfxsize=0.40\textwidth \epsfbox{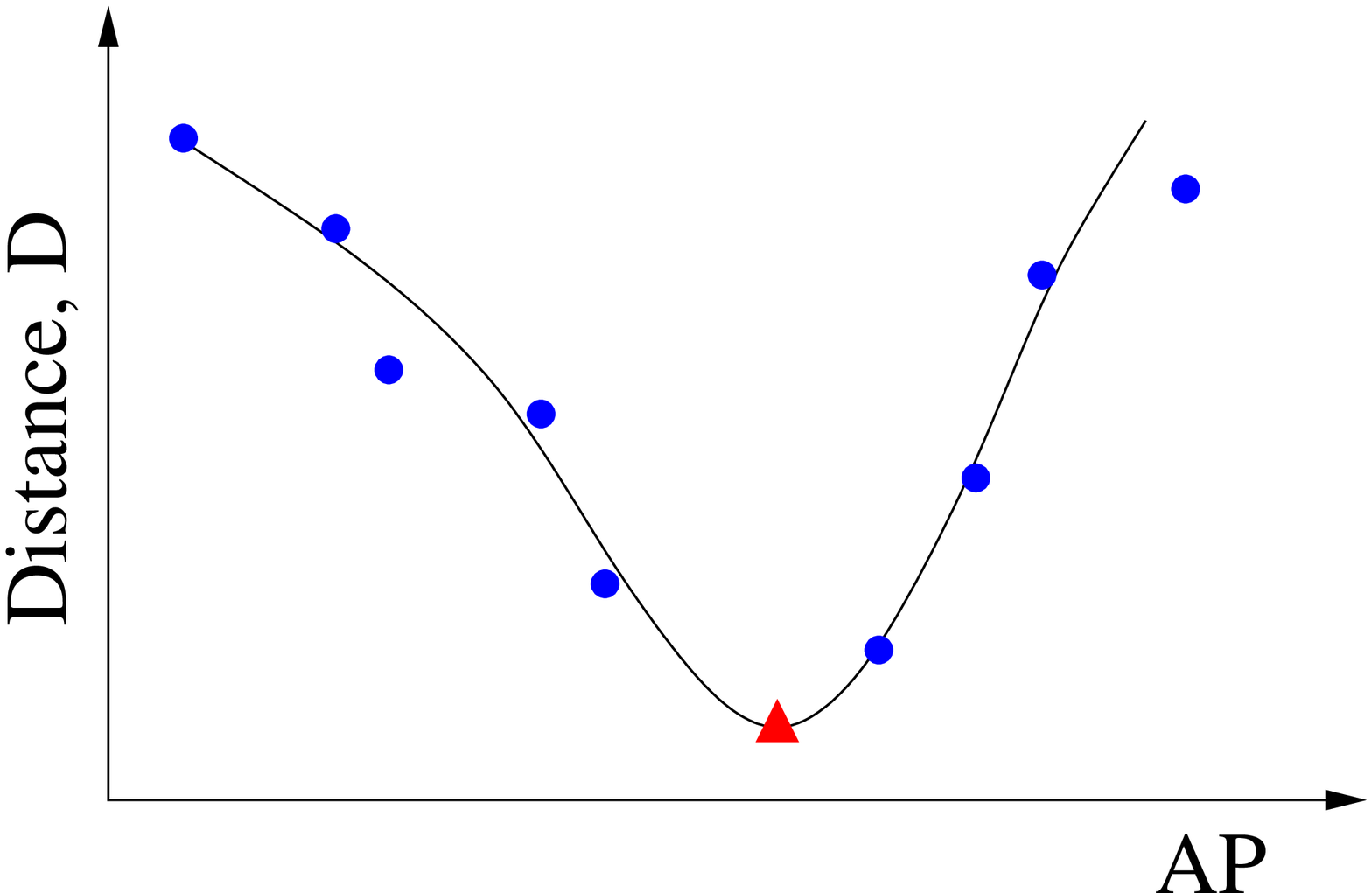}
\end{center}
\caption{Schematic illustration of the generic minimum distance method
(MDM). Top: a two-dimensional data space populated with pre-classified
templates.  Assigning parameters to a new object involves
looking at the APs of the nearest neighbours (with the data dimensions
suitably scaled).  APs are assigned either by interpolating in the
data space (i.e.\ solving the function APs\,=\,f\,(data) locally at the
new object -- in the simplest case this is just an average of one or
more neighbours) or in the parameter space (i.e.\ minimising the function
D\,=\,g\,(APs), shown for one AP in the bottom panel).}
\label{mdm}
\end{figure}

Classification and parameter estimation is the problem of assigning
object classes or APs and generally involves determining some kind of
mapping from the data space to the parameter space
(Fig.~\ref{mapping}).\footnote{The {\it data space} refers to the data
acquired from Gaia, such as fluxes in different filters or the RVS
spectrum. The {\it parameter space} refers to those properties of the
sources we wish to determine, such as \teff\ or extinction, but could
also refer to discrete classes (e.g.\ star, galaxy, quasar).}  A
frequently used approach is the {\it supervised} or pattern matching
approach, in which pre-classified data ({\it templates}) are used to
infer the desired mapping. This mapping is then applied to new data to
establish their classes or APs. Perhaps the most familiar such
technique is the minimum distance method (MDM), shown schematically in
Fig.~\ref{mdm}. This is a {\it local} template matching method, in
which only the properties of the local neighbours in the data space
influence the APs of the new object.  Here, we make a local fit of the
mapping function, or even just assign the APs of the nearest template
to the new object. 

However, such local approaches quickly run into the well-known `curse
of dimensionality'.  As the dimensionality, $p$, of the data space
increases, the density of the templates decreases (for a fixed number
of templates).  For example, with MDM we may assign APs by averaging
the APs of those nearest neighbours selected such that they fill a
fraction $x$ of the entire data space around the new object. To do this
in $p$ dimensions, and if the templates were uniformly spaced, we
would have to include neighbours out to a fractional distance of order
$d = x^{(1/p)}$ in each data dimension.  With $x=0.01$ and $p=2$ this
gives $d=0.1$, i.e.\ templates out to 10\% of the full range of each
data dimension are included. But if we increase the number of data
dimensions (i.e.\ if we have more filters or more spectral bins),
to, say $p=10$, then we see that $d=0.63$. That is, the ``nearest''
neighbours now extend to 63\% of the entire range of each data
dimension. Such distant neighbours are likely to be completely
unrelated to the new object. The result is that we get a very large
bias in our AP estimation using local methods.  The only way to avoid
this is if we increase the number of templates exponentially with $p$,
but this quickly becomes inhibitive. If we simply shrink our
neighbourhood volume we may not have any neighbours at all, or, if we have
only one or two neighbours then we will get a large variance in the AP
estimate.\footnote{For more on the curse of dimensionality and the
bias--variance trade-off, see, for example, Hastie et al.\ 2001.}

A more sensible way of overcoming this problem is to either use
structured regression or a global regression approach.  In the former
we compensate for the lack of data at a local scale by making
assumptions about the shape or properties of the mapping function.
With global regression we also do this, but we furthermore use all of
the available data to form a single regression over the entire data.
Examples of this include neural networks and support vector machines.

\begin{figure}[t]
\begin{center}
\epsfxsize=0.35\textwidth \epsfbox{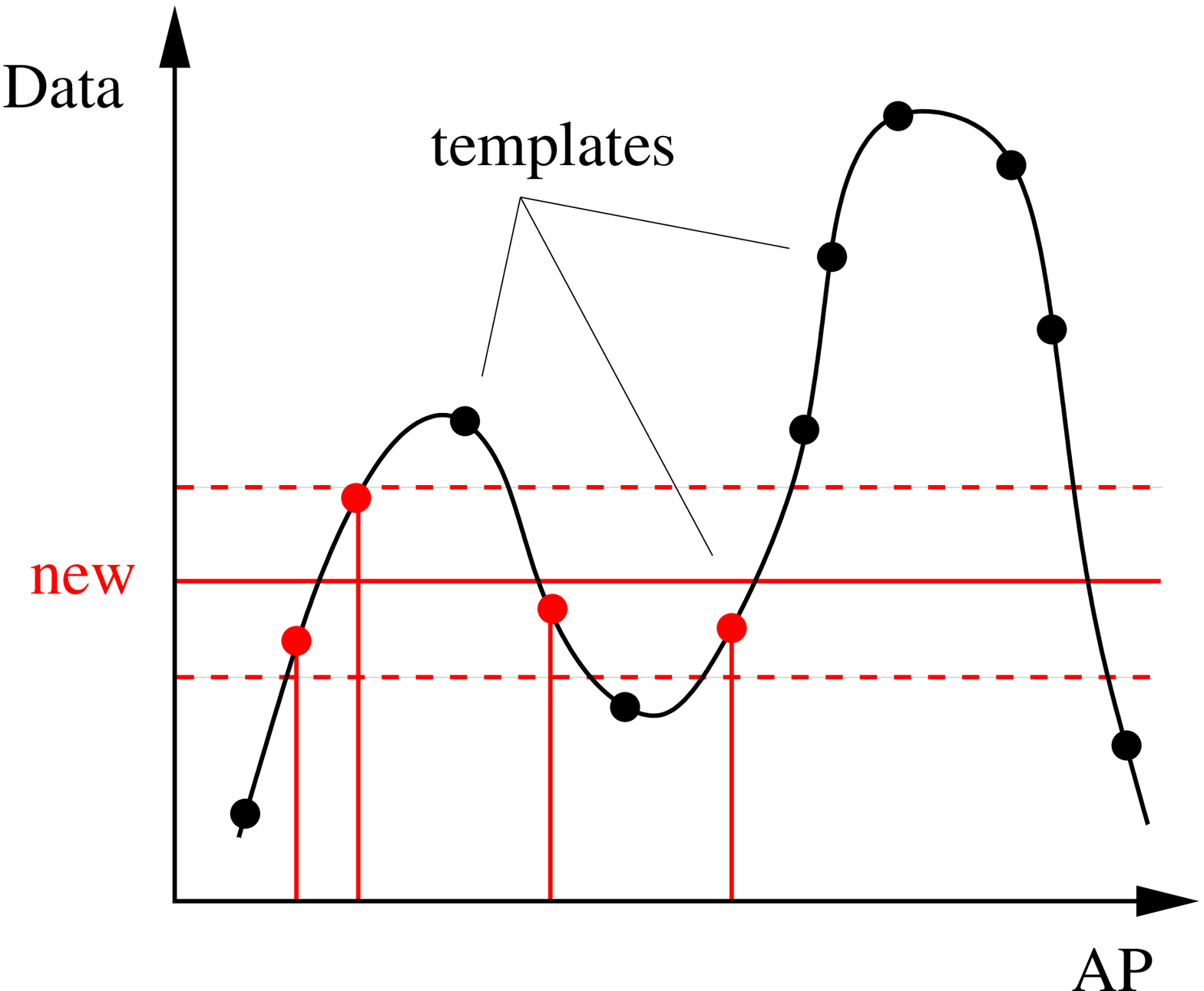} 
\epsfxsize=0.35\textwidth \epsfbox{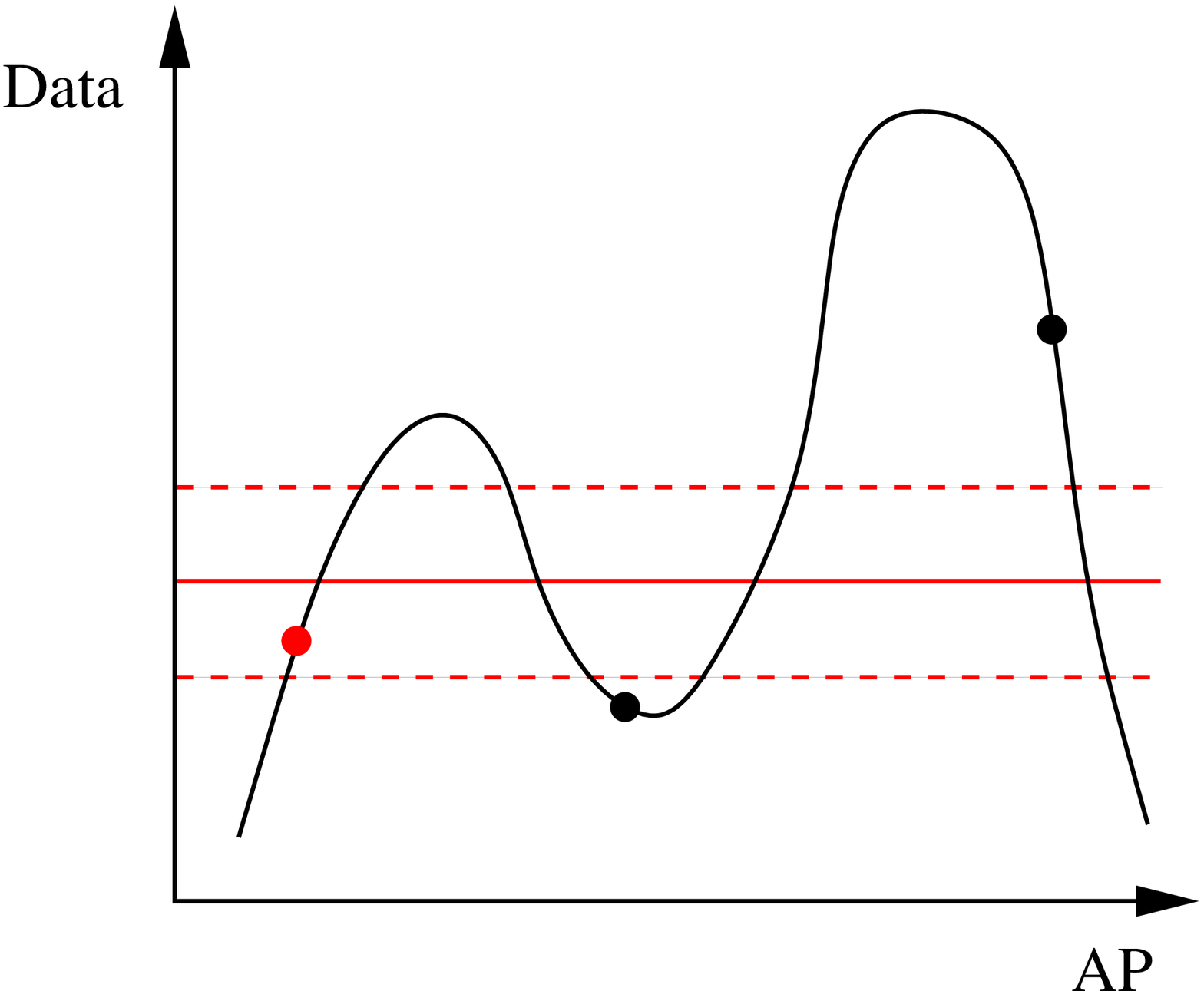}
\end{center}
\caption{Schematic diagram of the functional relationship between a
one-dimensional data and a one-dimensional AP space showing
degeneracies. We see degeneracies, i.e.\ several AP solutions for a
single given data measurement (shown by the horizontal line).  The
dashed lines show the noise level, so that any of the four templates
in the top diagram consistent with this (shown with vertical lines)
are consistent with this measurement.  We cannot choose between them.
If we had a lower density template grid, we would even miss entire
parts of the AP space as solutions.}
\label{degeneracy}
\end{figure}

There are, however, additional issues.  Classification and AP
estimation is the process of mapping from the data space to the AP or
class space.  By contrast, simulation of source SEDs is the opposite
mapping, i.e.\ from the AP space to the data space.  The fomer mapping
which we are interested in determining is therefore an inverse
mapping.  This is generally non-unique. In other words, whereas a
given set of APs provides a unique SED, two sets of APs could produce
the same SED. This is compounded by the effects of noise. The larger
the photometric noise on a SED or set of flux measurements, the greater
the number of different possible sets of APs this could correspond to.

This is illustrated schematically in Fig.~\ref{degeneracy}. In the
top panel we see that there are four templates (those lying within
the noise bounds) which give rise to data consistent with the new
observation. Confronted with this degeneracy we must decide what to
do. Do we quote all results? Do we average the APs?  There are in fact
whole ranges of the AP which are consistent with the data, so an
unweighted average will be biased by the distribution of the nearest
templates.  Moreover, at large AP, there is actually another solution
which we have completely failed to recognise due to the low density of
templates in that region.  The problem is worse with a lower density
template grid (bottom panel), or, equivalently, lower noise
data. Clearly, a local method which just assigns the APs using the
nearest neighbours will give biased results.

But even with a global regression we have problems, as the function we
are trying to approximate may not be single valued, so the regression
could go very wrong where we have these degeneracies.  (Think of
rotating the top panel of Fig.~\ref{degeneracy} by 90$^{\circ}$ and
trying to fit a single-valued function through the templates.)

\begin{figure*}[t]
\begin{center}
\psfig{width=0.65\textwidth, angle=90, figure=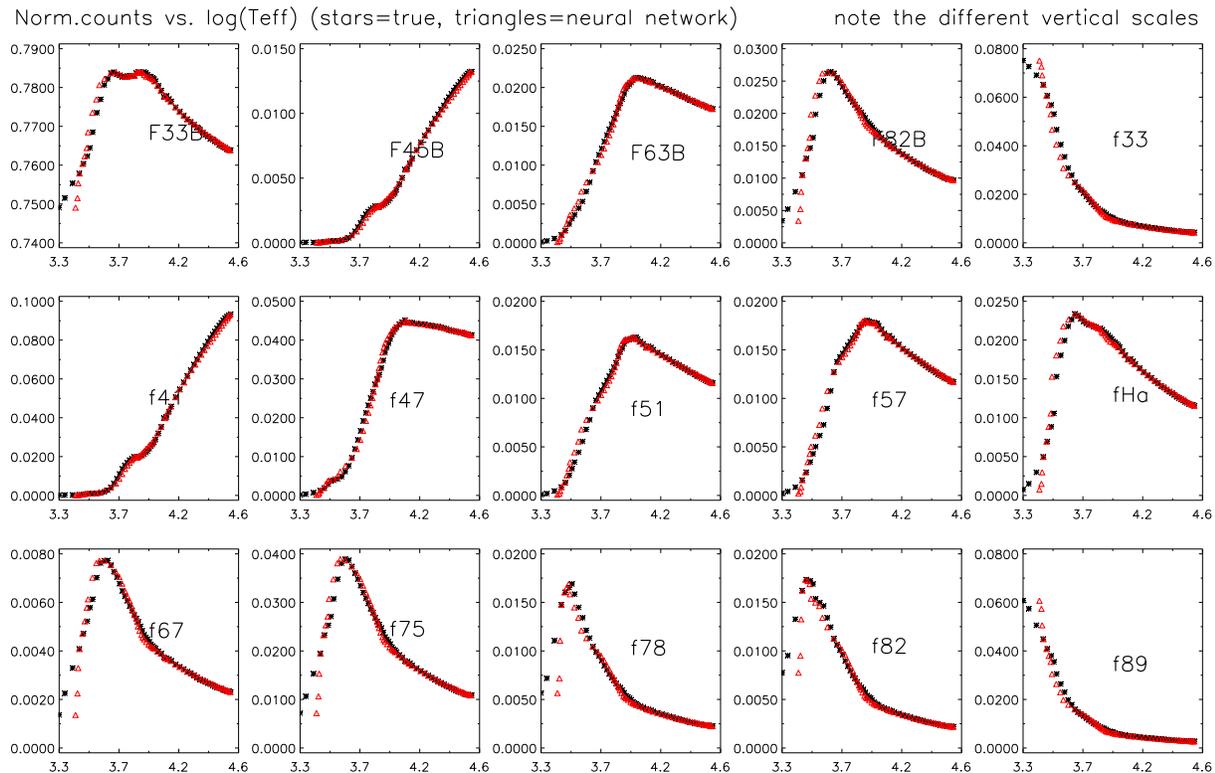} 
\end{center}
\caption{Each panel shows show the photon counts in 
a filter varies as a function of log\,\teff\ for stars with
\logg\,=\,4.0 and \feh\,=\,0.0 and zero extinction. The filter system
is the 1X BBP+MBP system, with each filter named according to the
first two digit of its central wavelength in nm. The trailing `B'
denotes a broad band (BBP) filter. The photon counts are normalized,
that is their sum across all filters for a star is the same for all stars.
The
stars show the simulated data (stellar spectra passed through
the Gaia instrument model) and the triangles the neural network
predictions.}
\label{pred_teff}
\end{figure*}

\begin{figure*}[t]
\begin{center}
\psfig{width=0.65\textwidth, angle=90, figure=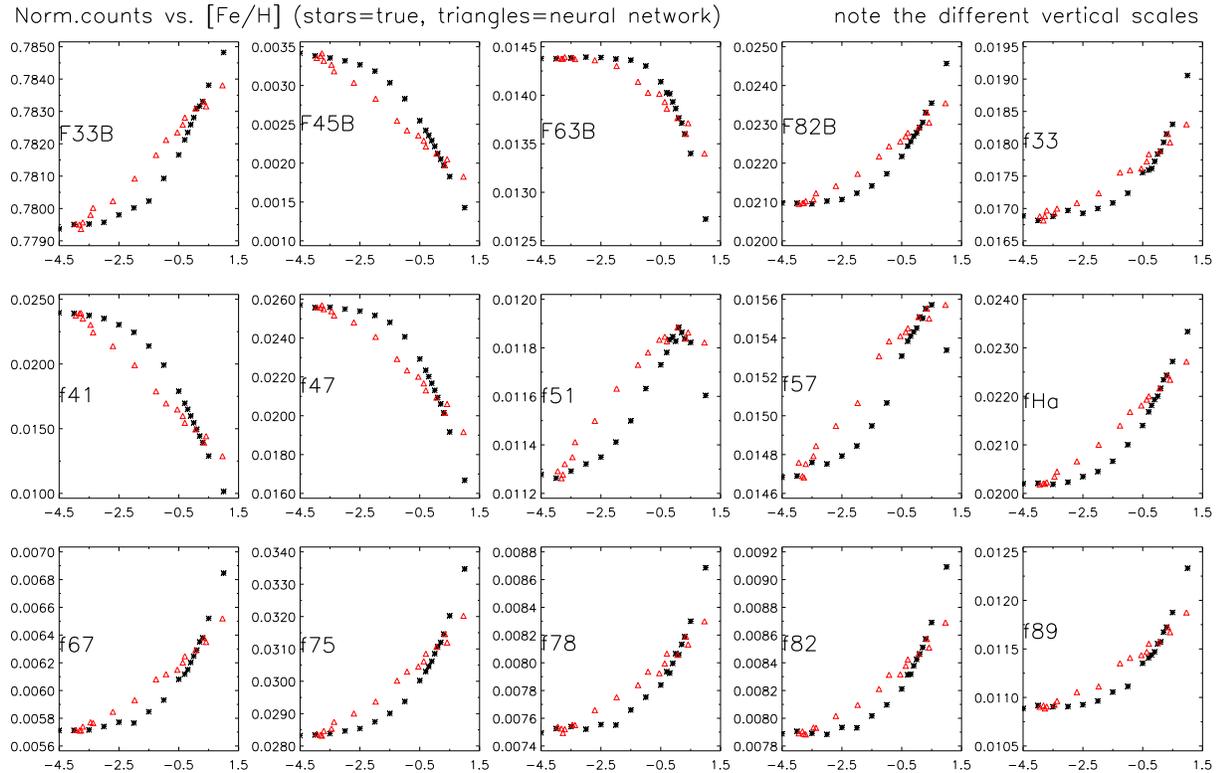} 
\end{center}
\caption{Same as Fig.~\ref{pred_teff}, but now showing photon counts as a function of \feh\ for fixed \teff\,=\,6000.}
\label{pred_feh}
\end{figure*}

We may assess this by training a global model (in this case a neural
network) on simulated Gaia photometric data and then examining the
mapping it has learned. This is shown in Figs.~\ref{pred_teff}
and~\ref{pred_feh}. Projections of the true relationship
between the filter fluxes and the APs are shown as stars. Projections
of the mapping learned by the network are shown as triangles. Of course, the
network is learning the mapping in the inverse sense from the way it
is plotted (that is, given the fluxes in all filters, it predicts the
APs). Both plots are derived from the same network which had 15 inputs
(one per filter) and 3 inputs (one per AP). The training data grid
consisted of many thousands of spectra over a wide range of \teff,
\logg\ and \feh\ combinations for stars of $G=15$ with simulated
end-of-mission photometry.

Examining Fig.~\ref{pred_teff} we see that the network has done very
well in determining the \teff\ mapping. It manages to reproduce most
of the small scale feature of the mappings. Only at the lowest
temperatures does it have problems. This is due at least in part to
the difficulty most regression models have at the boundaries of data
sets, which arises because the regression is essentially only constrained on
one side. In many of these one-dimensional cuts, the data to AP mapping 
is not single valued. The fact that the neural network
can nonetheless produce the correct mapping shows that in this 15
dimensional data space the stellar data must lie on a lower
dimensional manifold which does not show any serious degeneracies (for
\feh\ and \logg\ fixed).  Fig.~\ref{pred_feh} shows the mapping as
a function of metallicity keeping \teff\ and \logg\ fixed. We straight
away see that even though the true mapping is simpler, the neural
network has more problems reproducing it.  The reason for this can be seen when
we compare the vertical (photon count) scale of the two plots: the
variance in counts across the full range of \feh\ is much smaller than
the count variance across the full range of \teff. Putting it another
way, \feh\ is a `weak' AP compared to \teff, in that varying \feh\ by
X percent of its full range has a much smaller effect on the data
(SED) than does varying \teff\ by the same amount. The effects of
\feh\ are subtler and therefore harder to extract from the dominant
effect of \teff. This is compounded further by noise (which is small
here) and the effects of the other parameters.\footnote{\logg\ varies
in this data set, so the network had to try and learn its mapping
too. Adding additional parameters, in particular interstellar
extinction (which is a `strong' parameter), make this harder still.}

\section{Areas currently under\\investigation}

The task of designing the classification system for Gaia, and for
developing, testing and implementing the required algorithms, is the
task of the Gaia Classification Working Group.\footnote{also called
ICAP, for {\it Identification, Classification and Astrophysical
Parametrization}, for the slightly pedantic reason that
`classification' strictly only refers to placing objects into discrete
boxes.} A number of different tasks have been completed or have made
some progress. These can be found in detail on the working group web
site.\footnote{Currently http://www.mpia.de/GAIA/ although it can
always be found via the main Gaia website at ESA-ESTEC.} Here I just
provide a brief summary of the main tasks and provide the reference of
the relevant working group document (things like ICAP-CBJ-013). These
can be obtained from the ICAP web site or from Livelink.
\begin{itemize}
\item{Design and implementation of the overall classification system.
ICAP-CBJ-002/Bailer-Jones (2002), ICAP-CBJ-007, ICAP-CBJ-011.}
\item{A `blind testing' procedure to assess the performance of various
photometric systems and classification algorithms for performing
discrete classification and for estimating stellar
parameters. ICAP-AB-003 (collation of results), ICAP-PW-001 (detailed
analysis of MDM and neural network results for AP estimation),
ICAP-CH-001 (analysis of various statistical algorithms for discrete
classification and outlier detection).}
\item{Minimum distance and perturbation methods. ICAP-VM-001.}
\item{Identification and parametrization of unresolved binaries. ICAP-PW-003.}
\item{Design of the Gaia photometric systems by direct optimization using
evolutionary algorithms. ICAP-CBJ-013/Bailer-Jones (2004),
GAIA-CBJ-016.}
\item{The effect of CNO and $\alpha$ elements on the Gaia photometry. 
ICAP-GT-002.}
\item{Stellar parameter uncertainty estimates using bootstrapping neural 
networks. ICAP-PW-004.}
\item{Classification of QSOs and determination of their intrinsic parameters, 
including photometric redshift. Claeskens et al.\ (these
proceedings).}
\end{itemize}
Several articles in the present proceedings describe some of these and
other classification issues in more detail. In particular see the
contributions by Bailer-Jones, Carrasco et al., Claeskens et al.,
Girard \& Soubiran, Maiz-Apellaniz, Malyuto, Picaud et al.,
Recio-Blanco et al., Willemsen et al. and Zwitter et al.

\section{Issues and future work}

Work on the classification system for Gaia is still in its early
days. Given the heterogeneity of the Gaia data, plus the enormous
range of objects and APs which it must deal with, classifying the Gaia
data is not a simple task of applying some black box classifier to the
entire data set. Some of the key issues which need to be addressed
are as follows.
\begin{enumerate}
\item{A proper handling of AP degeneracies.  If degeneracies cannot be 
avoided in the photometric system (and they almost certainly cannot
be), then we must at least identify where they occur. Because we know
the forward mapping (data $\rightarrow$ APs) from stellar models, we
can in principle determine this. This information could be
used to perform a partition of the data space such that each partition
is handled by a separate regression model, free of degeneracies.}
\item{Coping with `weak' APs. Methods exist for boosting the sensitivity of
a regression model to weak APs, e.g.\ increasing their contribution in
the error minimization used to train the model (this is already used
in our neural network models). However, other approaches need to be
considered, such as iterative or hierarchical approaches in which
first the strong APs are determined and then the weaker ones given our
knowledge of the strong APs.}
\item{Dealing with systematic and correlated errors. 
This is related to the issue of `strong' and `weak' APs.}
\item{Training and regularization of models. While global models have 
some advantages over local ones, they are sensitive to the
distribution (across APs) of the training data. This can introduce
biases in the sense that it implicitly sets a prior probability on the
APs of new objects.}
\item{Combining heterogeneous data. For sufficiently bright objects we have 
available two sets of photometric data (at different spatial
resolutions), spectroscopic data and a parallax. It needs to be
carefully considered how, from an algorithmic perspective, these data
should best be combined to yield a self-consistent set of APs for a
object. On the other hand, discrepancies could be important as they
could indicate peculiar objects. Not all objects can be treated the
same way. For example, the quality of the RVS data will vary
enormously, and the good estimates of the data uncertainties which we
can get from our instrument models should also be utilized.}
\item{Calibration. So far we have trained models and assessed performance
mostly using synthetic data, as this is the only source of data at the
required resolution and wavelength coverage (for simulating the Gaia
photometry) covering the required wide range of APs.  Determining
physical parameters is the ultimate goal, so at some level stellar
models and synthetic spectra must be used. But there will be
limitations if we rely only on synthetic data for training models. One
idea is to use a limited set of actual Gaia data of well-known stars
to calibrate the classification models, for example by adjusting the
training data. If insufficient well-known stars (with well determined APs)
exist in Gaia's catalogue, additional ground-based observations
will be necessary to better characterize them in a homogeneous way. Such 
observing programs could and should start before the Gaia mission.}
\item{So far no attention has been paid to using morphological information, for
example to perform star/galaxy discrimination. This will be partly the
role of the onboard detection algorithms.}
\item{A number of issues have not yet been addressed, although work is 
starting on them. These include: galaxy classification; asteroid
classification; classification of special types of stars (i.e.\ those
for which we do not have reliable synthetic spectra); dealing with
anomalous extinction. Finally, some kind of unsupervised
classification, or internal classification scheme, should be
investigated. Such approaches are independent of any physical models
or pre-defined classification scheme and allow us to look for natural
groupings within the data. This type of exploratory data analysis is
an important complement for discovering new types of objects.}
\end{enumerate}


\end{document}